\documentclass[prd,twocolumn,nofootinbib,twocolumn]{revtex4-2}

\usepackage{graphicx, epsfig}
\usepackage{color}
\usepackage{mathrsfs}
\usepackage{bm}
\usepackage{amsmath,amssymb}% for \eqref
\usepackage{mathrsfs}
\usepackage[caption=false]{subfig}
\usepackage[normalem]{ulem}
\usepackage{mathtools}
\usepackage[x11names]{xcolor}
\usepackage{svg}

\usepackage{physics}
\usepackage{amsthm}
\usepackage{soul}
\usepackage[colorlinks = true, linkcolor = purple, urlcolor  = blue, citecolor = blue, anchorcolor = blue]{hyperref}
\usepackage{float}
\usepackage[caption=false]{subfig}
\graphicspath{{images/}}

\definecolor{fashionfuchsia}{rgb}{0.96, 0.0, 0.63}
\colorlet{no_so_fashion_purple}{blue!50!red}

\newcommand{\be}{\begin{equation}}
\newcommand{\ee}{\end{equation}}
\newcommand{\ba}{\begin{eqnarray}}
\newcommand{\ea}{\end{eqnarray}}

\newcommand{\nn}{\nonumber}

\newcommand{\la}{\langle}
\newcommand{\ra}{\rangle}

\newcommand{\mJ}{{\mathcal{J}}}
\newcommand{\mK}{{\mathcal{K}}}

\begin{document}

\title{Production of global vortices with quantum mediation}
\author{Omer Albayrak$^*$, Tanmay Vachaspati$^{*}$}
\affiliation{
$^*$Physics Department, Arizona State University, Tempe,  Arizona 85287, USA.
}

\begin{abstract}
We study global vortex production in (numerical) scattering experiments when the 
scatterer and the vortex degrees of freedom interact only due to intermediary quantum variables.
We work in $2+1$ dimensions with a complex scalar field, $\phi$, that supports global vortices, a real scalar field, $\psi$, that is the scatterer, and a quantum field, $\rho$, that couples to both $\phi$ and $\psi$, acting as a mediator.
We simulate the scattering of relativistic Gaussian wavepackets of $\psi$ and scan parameter space for regions where vortex-antivortex pairs are produced.
The results show that vortex production is highly sensitive to the initial Gaussian parameters, and the parameter space is chaotic with ``holes" and isolated regions of vortex production. 
\end{abstract}

\maketitle

\section{Introduction}
\label{intro}

The soliton sector of a quantum field theory is generally described by classical
solutions of the field equations with quantized fluctuations around the solution~\cite{Coleman:1985rnk,Rajaraman:1982is,Weinberg:2012pjx,Vachaspati:2006zz}. On the other hand, the particle sector is generally described in terms of perturbative 
quantum field theory. The transition from particles to solitons is the subject of this paper.

More specifically, we are envisioning a process similar to the creation of magnetic
monopoles by scattering light, {\it i.e.} a large number of photons. 
Both the initial state (light beams) and the final state (magnetic monopoles) are well described
in classical terms. However, the issue is that there is no classical path from light 
to monopoles because (classically) light does not interact with light~\cite{Jackson:1998nia}. Thus, light
on light scattering can only produce magnetic monopoles if we include some
intermediate quantum interactions~\cite{Euler:1935zz,Heisenberg:1936nmg,PhysRev.46.1087}. This study is similar in spirit to the creation of solitons in purely classical scattering as in Refs.~\cite{Dutta:2008jt,PhysRevLett.105.081601,Demidov:2011eu,Vachaspati:2011ad,PhysRevD.87.065018,Vachaspati:2016abz,Saurabh:2018bes,Saurabh:2019rrp}.

The creation of magnetic monopoles from light is a problem in three spatial dimensions,
and in an $N^3$ simulation volume with quantum mediation taken into account,
scales as $N^6$~\cite{Vachaspati:2018hcu} and is difficult to simulate. Instead, we focus on the problem in
two spatial dimensions in which the scattering of classical waves of a scalar field
result in the production of global vortex-antivortex pairs~\cite{Vilenkin:2000jqa,Manton:2004tk,Gibbons:1990gp}, but mediated by a second 
quantum scalar field. 
Even with the simplification of scattering in 2+1 dimensions, we encounter new
complexities. The first issue is technical, having to map a four-index variable to
a two-index variable, as explained in Sec.~\ref{model}. The second issue is more subtle, though less technical. In our earlier work~\cite{Albayrak:2023dul}, the
classical scattering of scalars followed by quantum mediation created kink-antikink 
pairs. If we apply the same idea in the context of vortices, we find that the dynamics
does not explore all of field space and, in particular, does not lead to the winding
in field space that is necessary for vortex production. To circumvent this issue, we
allow for small deviations of the vortex fields away from zero, as explained in
Sec.~\ref{initial}.

In Sec.~\ref{model} we describe the field theory we choose and our analysis
technique. In Sec.~\ref{initial} we set up initial conditions for the scattering, followed
by a description of numerical methods in Sec.~\ref{numerical}. Our results are
described in Sec.~\ref{results}, and we conclude in Sec.~\ref{conclusions}.

\section{Model}
\label{model}

We consider a complex scalar field $\phi$ that will be treated classically and has
global vortex solutions, a second scalar field $\psi$ that will also be treated classically,
and a third scalar field $\rho$ that is treated quantumly and interacts with both
$\phi$ and $\psi$. There are no direct interaction terms between $\phi$ and $\psi$.
We will choose ``large'' initial conditions in the $\psi$ field that will be treated
classically. The only channel to produce vortices is first through the production of
$\rho$ that can then produce vortices.
The Lagrangian is,
\begin{eqnarray}
L &=&  \partial_\mu \phi^* \partial_\mu \phi 
+ \frac{1}{2} (\partial_\mu \psi )^2
+ \frac{1}{2} (\partial_\mu \rho )^2 
- \frac{\lambda}{2} (|\phi|^2 - \eta^2)^2
\nn \\ 
&-& \frac{m_\psi^2}{2} \psi^2
- \frac{1}{2} \left ( m_\rho^2 + \alpha |\phi|^2 + \beta \psi^2 \right ) \rho^2
\label{eqn:the_lagrangian}
\end{eqnarray}
%where $\phi$ is a complex field and $\rho$ and $\psi$ are considered to be real fields.
and the equations of motion are,
\begin{eqnarray}
    \Box\phi  + \lambda \phi (|\phi|^2 -\eta^2) +\frac{\alpha}{2} {\rho}^2\phi &=& 0 
    \label{eqn:phiBefore}\\
    \Box\psi  + \left ( m^{2}_{\psi} +\beta {\rho}^2 \right )\psi &=& 0 \label{eqn:psiBefore} \\
     \Box\rho  +  \left ( m_\rho^2 + \alpha |\phi|^2 + \beta \psi^2 \right ) \rho &=& 0. \label{eqn:rhoBefore}
\end{eqnarray}

The quantum operator $\rho$ appears in the classical equations of motion for $\phi$ and
$\psi$. We will use the semiclassical approximation and replace $\rho^2$ by its expectation
value in its 
instantaneous quantum state, $\la \rho^2 \ra$, in those equations.
The equations for the classically treated fields are then given by,
\begin{eqnarray}
    \Box\phi  + \lambda \phi (|\phi|^2 -\eta^2) +\frac{\alpha}{2} \la {\rho}^2 \ra \phi &=& 0 
    \label{eqn:phiSC}\\
    \Box\psi  + \left ( m^{2}_{\psi} +\beta \la {\rho}^2 \ra \right )\psi &=& 0 \label{eqn:psiSC}
\end{eqnarray}
The dynamics of the quantum operator $\rho$ is 
described using the ``classical-quantum correspondence'' (CQC) as we now elaborate~\cite{Vachaspati:2018llo,Vachaspati:2018hcu,Vachaspati:2018pps}.

The CQC of $\rho$ follows a similar procedure as in~\cite{Albayrak:2023dul} with a slight change to accommodate the extra dimension, as we demonstrate below. The action for $\rho$ in 2+1 dimensional spacetime is written by,
\begin{equation}
    \mathcal{S}_{\rho} = \int d^3x \frac{1}{2} \Bigg[ \dot \rho^2 - (\nabla\rho)^2 - (m_{\rho}^2 + \alpha|\phi|^2 + \beta\psi^2 ) \rho^2   \Bigg] \label{eqn:rho_action_cts}
\end{equation}
where the dot denotes the time derivative.
The first step of the procedure is to discretize the action. We consider an $N \cross N$ lattice with lattice spacing 
$\Delta x=\Delta y = a$ 
in $x$ and $y$ axes respectively. 
Then, any field and its spatial derivative are defined as,
\begin{equation}
    f(t,x,y) \rightarrow f(t, ja, ka) \equiv f_{jk}(t) \label{eqn:discretize_field}
\end{equation}
\begin{align}
    \nabla^2 f_{jk}(t) =  \frac{1}{a^2} ( f_{j+1k}(t) - 2f_{jk}(t)+ f_{j-1k}(t) ) \nn \\
    + \frac{1}{a^2} ( f_{jk+1}(t) - 2f_{jk}(t) + f_{jk-1}(t) ) \label{eqn:discretize_derivative}
\end{align}
where  $j,k = 1,2,...,N$. Along with the periodic boundary conditions, the action then reads,
\begin{equation}
    \mathcal{S}_\rho = \int dt \frac{1}{2} dx dy \sum_{j,k} \{ \dot \rho_{jk}^2  - \rho_{jk} \Omega^2_{jl,km} \rho_{lm} \} \label{eqn:rho_action_disc}
\end{equation}
where $\Omega^2_{jl,km}$ is a 4-index object, but is not a tensor that is defined as,
\begin{eqnarray}
    \Omega^2_{jl,km} &=& \tilde \delta_{jl}\tilde \delta_{km} 
    \Bigg [ \frac{4}{a^2} +( m_{\rho}^2 + \alpha |\phi_{jk}|^2 + \beta \psi_{jk}^2 ) \Bigg ] \nn \\
    &-& 
    \frac{1}{a^2}  \left ( \tilde \delta_{j+1,l}\tilde \delta_{km}  + \tilde \delta_{j-1,l}\tilde \delta_{km} \right ) \nn \\
    &-& 
    \frac{1}{a^2} \left ( \tilde \delta_{jl}\tilde \delta_{k+1,m}  + \tilde \delta_{jl}\tilde \delta_{k-1,m} \right )   \label{eqn:omega4index}
\end{eqnarray}
where  $\tilde \delta_{jk}$'s are Kronecker deltas with periodic boundary conditions built-in, i.e., for $j=N$, $\tilde \delta_{l,j+1} = \delta_{l1}$.

\begin{figure}
    \centering
    \includegraphics[width=1.0\linewidth]{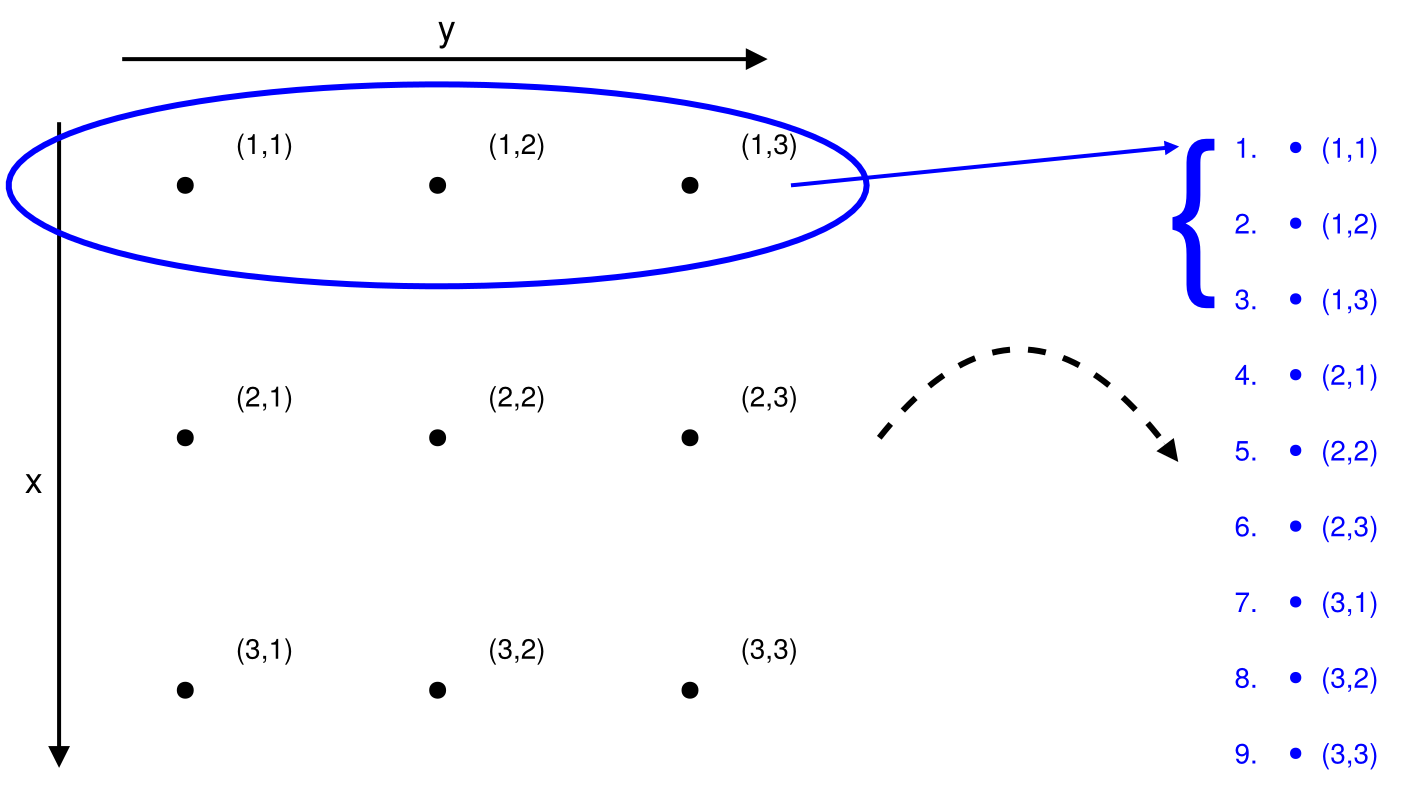}
    \caption{$3\cross 3$ example of flattening 2d lattice into 1d}
    \label{fig:flatten}
\end{figure}

This form of the action is inconvenient because
$\Omega^2_{jl,km}$ is a 4-index object, whereas it is easier
to work with ordinary matrices.
To get back the functionality of the matrix notation, we map the 2-dimensional lattice to a flattened 1-dimensional one. This way, the 4-index $\Omega^2_{jl,km}$ can be written as a 2-index matrix. Fig. \ref{fig:flatten} shows how this operation is conducted on a simple lattice. The map is given by,
\begin{equation}
    \mathcal{J} = (j-1)N +k \label{eqn:flatten_map}
\end{equation}
where  $j$ and $k$ represent the $x$ and $y$ lattice coordinates, respectively, $N$ is the size of the lattice and $\mathcal{J}=1,...,N^2$. Under this mapping, any field on our lattice will appear as a vector of size $N^2$,
\begin{equation}
    f_{jk} \rightarrow f_{\mathcal{J}}. \label{eqn:flatten_field}
\end{equation}
The periodic boundary conditions of the 2-dimensional lattice need to be carefully translated to the new flattened lattice. 
Starting from the beginning, each block of $N$ points has a constant $x$-coordinate with each element in the block being a different $y$-coordinate. And each block itself corresponds to a different $x$-coordinate. The consecutive lattice points in each direction are then given as follows
\begin{eqnarray}
    (x + dx, y) &=& (j+1,k) \rightarrow \mathcal{J}+N \nn \\
    (x, y + dy) &=& (j,k+1) \rightarrow \mathcal{J}+1.  \label{eqn:flatten_coord}
\end{eqnarray}
Then, the periodic conditions can be written as follows; $\forall k$, in the case of $j=N$, 
\begin{equation}
    (j+1,k) \rightarrow \mathcal{J}+N \pmod N = \mathcal{J} - N(N-1) \label{eqn:x_pbc_flattened}
\end{equation}
and $\forall j$, in the case of $k=N$, 
\begin{equation}
    (j,k+1) \rightarrow \mathcal{J} - (N-1). \label{eqn:y_pbc_flattened}
\end{equation}
Similar expressions can be written going the opposite direction for each coordinate. These can be easily seen in Fig. \ref{fig:flatten} for $N=3$.

With this mapping in mind, we can write the action for $\rho$ as,
\begin{equation}
        \mathcal{S}_\rho 
        = \int dt \frac{1}{2} dx dy \Big[ \dot{\boldsymbol{\rho}}^2  - \boldsymbol{\rho}^T \boldsymbol{\Omega^2} \boldsymbol{\rho} \Big] \label{eqn:rho_action_discrete}
\end{equation}
where 
$\boldsymbol{\rho} = (\rho_1, ...,\rho_{N^2})$ and 
$\boldsymbol{\Omega^2}$ is an $N^2 \cross N^2$ matrix which is the flattened 4-index $\Omega^2_{jl,km}$ from \eqref{eqn:omega4index}. Now that we have the action written compactly in a matrix notation, we can start the standard CQC procedure.
The Hamiltonian corresponding to the action $\mathcal{S}_\rho$ is then given by,
\begin{equation}
    H_\rho = \frac{1}{2}  \Bigg[ \frac{1}{a^2}  \mathbf{p}^T \mathbf{p} +  a^2 \boldsymbol{\rho}^T \mathbf{\Omega}^2 \boldsymbol{\rho} \Bigg] \label{eqn:rho_hamiltonian}
\end{equation}
where $\mathbf{p} = a^2 \dot{\boldsymbol{\rho}}$.
This Hamiltonian of $N^2$ coupled quantum harmonic oscillators can be mapped to coupled $N^4$ classical ones using Bogoliubov transformations~\cite{Vachaspati:2018hcu}. The dynamical variables of the resulting system are written as the matrix $\mathbf{Z}(t) = [Z_{\mathcal{JK}}(t)]$ with the corresponding conjugate momentum $\mathbf{P}(t) = [P_{\mathcal{JK}}(t)]$. The mapping is given by
\begin{align}
    \boldsymbol{\rho} =\frac{1}{a} &\Big[  \mathbf{Z^*}\mathbf{a}_0 + \mathbf{Z}\mathbf{a}_0^{\dagger T}  \Big]\label{eqn:rhotoZ}\\
    \mathbf{p}        =a           &\Big[  \mathbf{P^*}\mathbf{a}_0 + \mathbf{P}\mathbf{a}_0^{\dagger T}  \Big]\label{eqn:ptoZ}
\end{align}
where $\mathbf{a_0}= (a_1, ..., a_{N^2})^T$ and $\mathbf{a_0^{\dagger}}= (a_1^{\dagger}, ..., a_{N^2}^{\dagger})$ are the stack of creation and annihilation operators of each quantum harmonic oscillator at the initial time.
Thus, the quantum field $\boldsymbol{\rho}(t)$ can be fully written as the classical field $\mathbf{Z}$ and $\mathbf{P} = \dot{\mathbf{Z}}$. The corresponding classical system has the following action,
\begin{equation}
    \mathcal{S}_c = \int dt \frac{1}{2} \textrm{Tr}[\dot{\mathbf{Z}}^\dagger \dot{\mathbf{Z}} - \mathbf{Z}^\dagger \mathbf{\Omega}^2\mathbf{Z}] \label{eqn:Z_action}
\end{equation}
with the following equations of motion,
\begin{equation}
    \ddot{\mathbf{Z}} + \mathbf{\Omega}^2 \mathbf{Z} = 0 \label{eqn:Z_eom}
\end{equation}
and the initial conditions,
\begin{equation}
    \dot{\mathbf{Z}}_0 = \frac{1}{\sqrt{2}} \sqrt{\mathbf{\Omega_0}} \quad \textrm{and} \quad \mathbf{Z}_0 = -\frac{\mathrm{i}}{\sqrt{2}} \sqrt{\mathbf{\Omega_0}}^{-1}. \label{eqn:ic_Z}
\end{equation}
The CQC provides an exact correspondence of the quantum system to its classical counterpart. Given this, to characterize the dynamics of the quantum field, only equation \eqref{eqn:Z_eom} and the initial conditions \eqref{eqn:ic_Z} are needed. Furthermore, the backreaction term $\la\rho^2\ra$ in the equations of motion of $\phi$ and $\psi$ can be written  in terms of its classical counterpart $\mathbf{Z}$ as,
\begin{equation}
    \la \rho_{\mathcal{J}}^2 \ra = \frac{1}{a^2} \sum_{\mathcal{K}=1}^{N^2} \mathbf{Z}_{\mathcal{JK}}^* \mathbf{Z_{\mathcal{JK}}}. \label{eqn:rho_vac_exp}
\end{equation}

One can see that in the limit of $N \rightarrow \infty$, \eqref{eqn:rho_vac_exp} has a $\log(N)$ divergence. This can be remedied by subtracting the fluctuations in the trivial vacuum,
\begin{equation}
        \la \rho_{\mathcal{J}}^2 \ra \rightarrow \la \rho_{\mathcal{J}}^2 \ra - \la \rho_{\mathcal{J}}^2 \ra_0
\end{equation}
where, 
\begin{equation}
    \la \rho_{\mathcal{J}}^2 \ra{_0} \equiv \frac{1}{a^2} \sum_{\mathcal{K}=1}^{N^2} \mathbf{Z}_{\mathcal{JK}}^* \mathbf{Z_{\mathcal{JK}}}\biggr |_0
\end{equation}
is the trivial vacuum expectation value with $\abs{\phi} = \eta$ and $\psi=0$.

By CQC, the energy of quantum field $\rho$ can be represented in terms of $\mathbf{Z}$,
\begin{equation}
    E_\rho =  \frac{1}{2} \Tr[\mathbf{P}^\dagger \mathbf{P} + \mathbf{Z}^\dagger \mathbf{\Omega^2} \mathbf{Z}]
\end{equation}
and the energy density is,
\begin{align}
    \epsilon_{\rho,i} = \frac{1}{2a^2} \sum_{\mathcal{K}=1}^{N^2} &\bigg\{ \abs{\dot{Z}_{\mathcal{JK}}}^2 + \abs{\mathbf{\nabla}{Z}_{\mathcal{JK}}}^2 \nn \\
    + &\bigg[ m_\rho^2 + \alpha\abs{\phi_{\mathcal{J}}}^2 + \beta \psi_{\mathcal{J}}^2 \bigg] \abs{Z_{\mathcal{JK}}}^2
    \bigg\} \label{eqn:edensity_rho}
\end{align}
where
\begin{align}
   \nabla Z_{\mathcal{JK}} &= \frac{1}{2a} (Z_{\mathcal{J}+1\mathcal{K}} - Z_{\mathcal{J}-1\mathcal{K}}  ) \nn\\ 
                            &+ \frac{1}{2a} (Z_{\mathcal{J}\mathcal{K}+1} - Z_{\mathcal{J}\mathcal{K}-1}  ).
\end{align}
All of the terms in \eqref{eqn:edensity_rho} come with the same $\log(N)$ divergences as above. To renormalize these, we use the same strategy as before with an additional zero-point energy subtraction,
\begin{equation}
    \epsilon_{\rho,J}^R = \epsilon_{\rho,J} - \frac{1}{2} \bigg[\alpha\abs{\phi_{\mathcal{J}}}^2 + \beta \psi_{\mathcal{J}}^2 \bigg] \la \rho_{\mathcal{J}}^2 \ra_0 - \epsilon_{\rho,\mathcal{J}} \big|_0. \label{eqn:edensity_renorm}
\end{equation}
In \eqref{eqn:edensity_renorm}, the last term is the zero-point energy density that is, the energy density of the field $\rho$ in the background of the trivial vacuum; $\abs{\phi}=\eta$ and $\psi=0$. 
Note that $\epsilon_{\rho,J}^R$ can be negative since
the vacuum energy subtraction can be larger than that of quantum fluctuations in the non-trivial $\psi$ background.
This effect is clearly observed at early times in Fig.~\ref{fig:s1-indv-energies} (Sec.~\ref{results}).

Then, the total energy in the field $\rho$ can be written as,
\begin{equation}
    E_{\rho}^R = E_\rho - \sum_{\mathcal{J}=1}^{N^2}  \frac{a^2}{2} \bigg[\alpha\abs{\phi_{\mathcal{J}}}^2 + \beta \psi_{\mathcal{J}}^2 \bigg] \la \rho_{\mathcal{J}}^2 \ra_0 - E_\rho \big|_0.
\end{equation}
Finally, the total energy of the system is,
\begin{equation}
    E = E_\phi + E_\psi + E_\rho^R
\end{equation}
where
\begin{align}
    E_\phi &= a^2 \sum_{\mathcal{J}=1}^{N^2} \bigg[ \abs{\dot{\phi}_{\mathcal{J}}}^2 + \abs{\nabla \phi_{\mathcal{J}}}^2 + \frac{\lambda}{2} (\abs{\phi_{\mathcal{J}}}^2 - \eta^2)^2 \bigg]   \\
    E_\psi &= a^2 \sum_{\mathcal{J}=1}^{N^2} \frac{1}{2}\bigg[ \dot{\psi}_{\mathcal{J}}^2 + (\nabla{\psi_{\mathcal{J}}})^2 +  m_\psi^2\psi_{\mathcal{J}}^2 \bigg].
\end{align}

In summary, the final discrete equations of motion that we need to solve for the system are as follows,
\ba
&&
    {\ddot{\phi}_\mJ} - \nabla^2\phi_\mJ 
    + \lambda\phi_\mJ(\abs{\phi_\mJ}^2 - \eta^2) \nn \\
    && \hskip 0.1 cm
        + \frac{\alpha}{2a^2} \sum_{\mK=1}^{N^2} \biggr ( \mathbf{Z}^*_{\mJ\mK} \mathbf{Z}_{\mJ\mK} - \mathbf{Z}^*_{\mJ\mK} \mathbf{Z}_{\mJ\mK} \biggr|_0 \biggr) \phi_\mJ
        = 0 \label{eqn:phi_eom}
\ea
\ba
&&
    {\ddot{\psi}_\mJ} - \nabla^2\psi_\mJ
    + m_\psi^2 \psi_\mJ \nn \\
    && \hskip 0.1 cm
        + \frac{\beta}{a^2} \sum_{\mK=1}^{N^2} \biggr( \mathbf{Z}^*_{\mJ\mK} \mathbf{Z}_{\mJ\mK} - \mathbf{Z}^*_{\mJ\mK} \mathbf{Z}_{\mJ\mK} \biggr|_0 \biggr) \psi_\mJ 
        = 0 \label{eqn:psi_eom}
\ea
\begin{equation}
        {\ddot{Z}_{\mJ\mK}} + \Omega_{\mJ\mathcal{L}}^2 Z_{\mathcal{L}\mK} = 0 \label{eqn:Z_eom_disc}
\end{equation}
where the Laplacians are calculated with second order spatial differences as in \eqref{eqn:discretize_derivative}.

\section{Initial conditions}
\label{initial}

\begin{figure*}
% \centering
\subfloat[]{\includegraphics[width=0.45\textwidth]
        {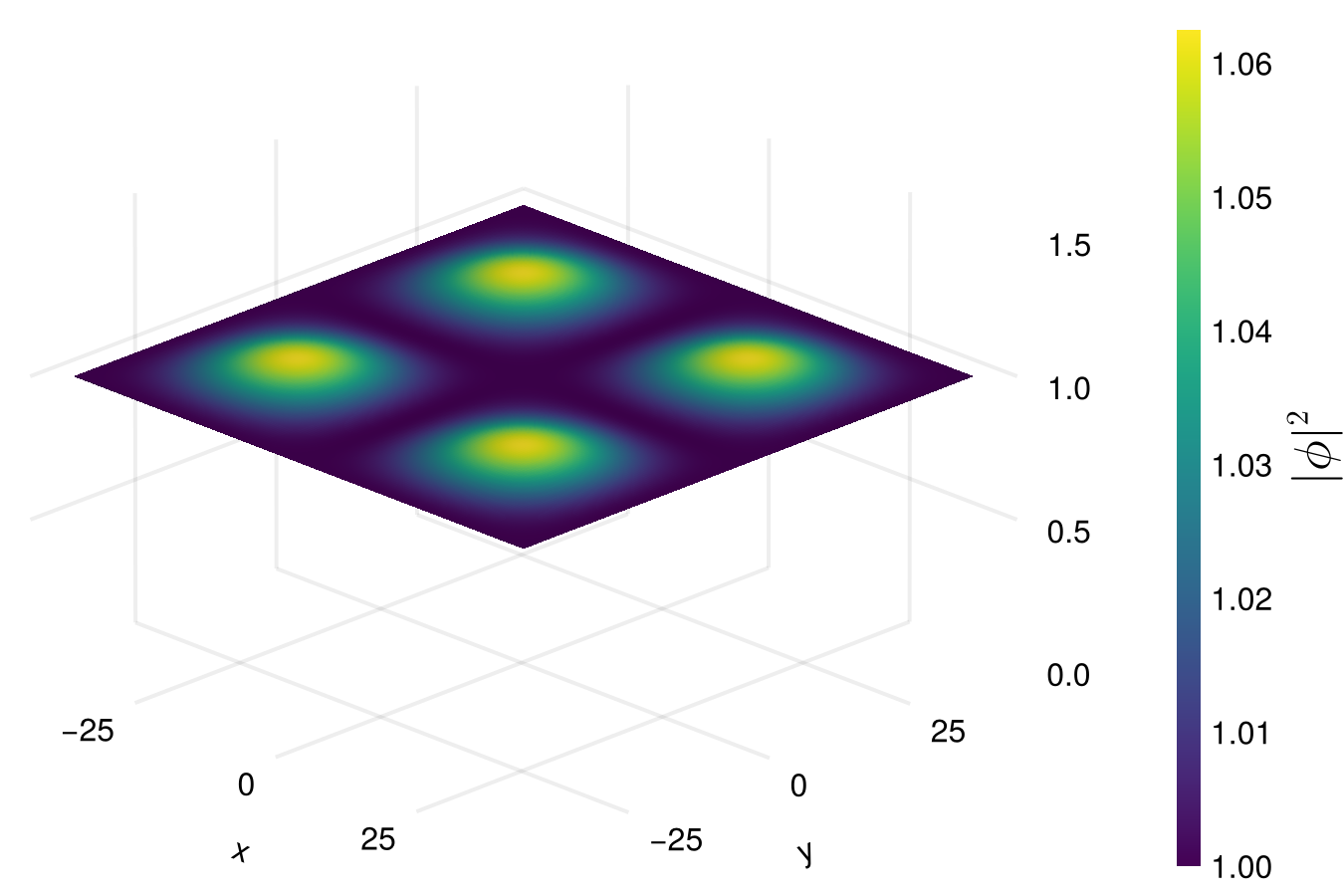}
        \label{fig:phi-ic}}
\subfloat[]{\includegraphics[width=0.45\textwidth]
        {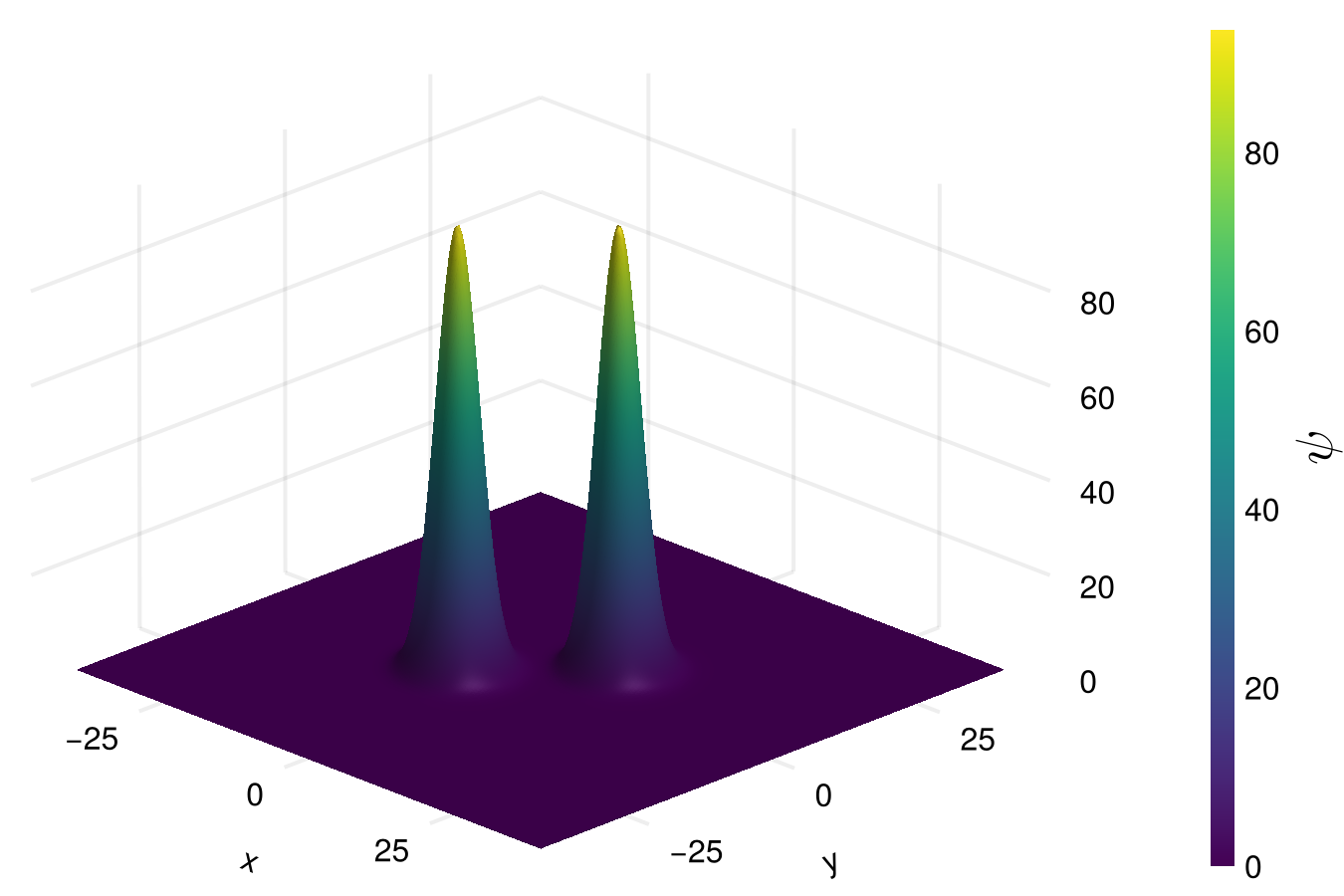}
        \label{fig:psi-ic}}
        \hfill
\caption{\label{fig:phi-psi-ic-main} Initial field configurations for $\phi$ and $\psi$
}
\end{figure*}

The main goal of this paper is to produce the simplest 2-dimensional topological defects, namely global strings (or vortices), using quantum interactions. 
In order to be specific, we consider the global $U(1)$ model,
\begin{equation}
    L_\phi = \partial_\mu \phi^*\partial_\mu\phi - \frac{\lambda}{2} (\abs{\phi}^2 - \eta^2 )^2
\end{equation}
where $\phi$ is a complex scalar field. $U(1)$ symmetry implies that we have a circle of degenerate vacuum states with $\abs{\phi} =\eta$. In other words, the vacuum manifold is $\mathbf{S}^1$. 

Since the only interaction of $\phi$ is with $\rho$, if vortices are to be produced, it would only be through quantum effects. To excite the quantum field, we consider scattering wavepackets for the field $\psi$ which solely interacts with $\rho$. The initial conditions for $\psi$ are chosen to be two Gaussian wavepackets that are relativistically moving towards each other with velocity $v=\sqrt{v_x^2+v_y^2}$ from a given initial separation $2r_0$ along the diagonal of the 2-dimensional lattice;
\begin{equation}
    \psi(t=0,x,y) = F(x_-,y_-) + F(x_+,y_+) \label{eq:ic_psi}
\end{equation}
where
\begin{equation}
    F(x,y) = Ae^{-kX(x,y)^2} e^{-kY(x,y)^2},
\end{equation}
$x_{\pm} = x\pm r_0$, $y_{\pm} = y\pm r_0$ and $X$ and $Y$ are the $x$ and $y$ coordinates boosted along the diagonal $x=y$ direction given by the Lorentz transformations at $t=0$;
\begin{align}
    X(x,y)\biggr|_{0} &= \Bigg[ 1 + \frac{(\gamma-1)v_x^2}{v_x^2+v_y^2}\Bigg]x 
                + \Bigg[\frac{(\gamma-1)v_xv_y}{v_x^2+v_y^2}\Bigg]y     \\ 
    Y(x,y)\biggr|_{0} &= \Bigg[1+ \frac{(\gamma-1)v_y^2}{v_x^2+v_y^2}\Bigg]y 
                + \Bigg[\frac{(\gamma-1)v_xv_y}{v_x^2+v_y^2}\Bigg]x
\end{align}
where $\gamma= 1/\sqrt{1-v^2}$ and the subscript $0$ refers to $t=0$. We also have,
\begin{align}
    \dot \psi(t=0,x,y) = \gamma v_x &\bigg [\partial_x F(x_-,y_-) - \partial_x F(x_+,y_+) \bigg] \nn \\
                       + \gamma v_y &\bigg [\partial_y F(x_-,y_-) - \partial_y F(x_+,y_+) \bigg] .
\end{align}

Initially, there are no vortices of $\phi$. 
If $\phi$ is chosen homogeneously to be in its vacuum state, the interaction term $\frac{\alpha}{2}\abs{\phi}^2\rho^2$ in \eqref{eqn:the_lagrangian} doesn't allow for production of any vortices. It only excites the radial modes of the potential for $\phi$. That is, the field would only move on a cross-sectional plane on the potential and wouldn't move along the angular direction. Vortices can only form when the field exhibits angular (phase) fluctuations that are strong enough to have a non-zero winding number, meaning that the phase changes by an integer multiple of $2\pi$ as one goes around a closed loop. Therefore, we've added small angular inhomogeneities to the initial field configuration. These fluctuations are given by,
\begin{equation}
    \delta\phi = i \, \mathcal{A} \sin(\kappa x) \sin(\kappa y)
\end{equation}
where $\mathcal{A}$ is a small scaling factor and $\kappa$ is chosen such that periodic boundary conditions are not violated;  $\kappa = \frac{2\pi}{L}$ where $L$ is the physical size of our lattice in any direction. Thus, the initial $\phi$ is given by;
\begin{equation}
    \phi(t=0,x,y) = \eta + \delta\phi(x,y), \ \ \dot\phi(t=0,x,y)=0 \label{eq:ic_phi}
\end{equation}

The quantum field $\rho$ is then chosen to be in its vacuum state in the background of  $\psi(t=0,x,y)$ and  $\phi(t=0,x,y)$ as in \eqref{eq:ic_psi} and \eqref{eq:ic_phi} respectively. 
This implies that the initial conditions in terms of $\mathbf{Z}$ and $\dot{\mathbf{Z}}$ 
are given by \eqref{eqn:ic_Z} such that $\mathbf{\Omega}_0^2$ is calculated with the initial conditions of $\phi$ and $\psi$. Fig. \ref{fig:phi-psi-ic-main} shows the initial $\phi$ and $\psi$ field configurations.

\section{Numerical method}
\label{numerical}

The lattice we choose to work with is a 2-dimensional $N \cross N$ periodic lattice with $N=200$ and lattice spacing $a=0.4$ in each direction. The time evolution of the system is conducted using the position Verlet method with the time step $dt=a/50$ for a light-crossing time. 

The physical parameters are chosen to be similar to~\cite{Albayrak:2023dul} to keep a consistent search for defects. We fix the couplings and masses as,
\be
\lambda = 1, \ \ \eta = 1, \ \  \alpha = 0.5, \ \ \beta = 0.5, \ \ m_{\rho} =1, \ \  m_\psi = 1.
\ee

The parameters we keep free are related to the initial conditions of the Gaussian wavepackets. We scan over the parameters $A$-amplitude, $v$-velocity, and $k$-width (see \eqref{eqn:scan_parameters}) while the initial separation $2r_0$ is chosen dynamically depending on these parameters for reasons that we explain below. We choose the boost velocity such that $v_x=v_y$ and thus $v=\sqrt{2}v_x$. The ranges for the parameters are,
\begin{equation}
    A \in [70,120], \ \ v \in [0.1,0.8], \ \ k=0.03,0.05,0.09  \label{eqn:scan_parameters}
\end{equation}
with $dA=1$, $dv=0.05$.

The Gaussian wavepackets disperse and don't retain their initial shape over time. Moreover, $A$, $v$, and $k$ all affect the initial spread of the wavepackets. Hence, to keep the search for vortices consistent, we adjust the initial separation $2r_0$ based on a fixed minimal overlap ($10^{-3}A$) between the wavepackets, for all given parameters, $A$, $v$, and $k$.

To detect the vortices, we employ a search algorithm that takes snapshots of the field configuration of $\phi$ at certain time intervals. We then calculate the winding number on each plaquette~\cite{Vilenkin:2000jqa};
\begin{equation}
    n = \frac{1}{2\pi} \oint d\theta  
\end{equation}
where $\theta$ is the phase of $\phi = \abs{\phi} e^{i\theta}$.
As we traverse a plaquette on the lattice, if the phase makes at least one complete rotation, namely $|n| \geq 1$, in field space - also known as ``winding'' - then the field must vanish at some point within the loop. 
This topological condition allows us to locate the vortices.
To identify individual vortices and distinguish them from overlapping regions or noise, we further impose that the regions with windings $|n| \geq 1$ be separated by a distance of at least $2/m_\phi$ where $m_\phi = \sqrt{2\lambda \eta}$ is the mass scale.

\section{Results}
\label{results}

The simulations demonstrate the creation of global vortices. Before discussing specific examples of vortex production, we note that the simulations are categorized by the maximum number of vortices generated during the entire simulation rather than the number that remain at the end. This choice is motivated by the strong attractive force between vortices and anti-vortices.
In the long-range interactions, the force between the vortices (and anti-vortices) is proportional to $F \propto - 1/R$, where $R$ is the distance to the nearest vortex, and the energy associated with a vortex field is given by~\cite{Vilenkin:2000jqa,Shifman:2012zz},
\begin{equation}
    E \approx 2\pi \eta^2 \ln{ \left( \frac{R}{\delta} \right) }
\end{equation}
where $\delta$ is the core size of a vortex. The energy required to separate the vortex-anti-vortex pairs created has a $\log$ divergence; they are strongly attracted to each other. Thus, the created pairs do not wander much and are annihilated fairly quickly in simulation times. Hence, we gauge the production efficiency with the maximum number of vortices and anti-vortices created during the simulation.

\begin{figure*}
    \centering
    \subfloat[]{%
    \begin{minipage}{0.315\textwidth}
        \centering
        \includegraphics[width=\linewidth]{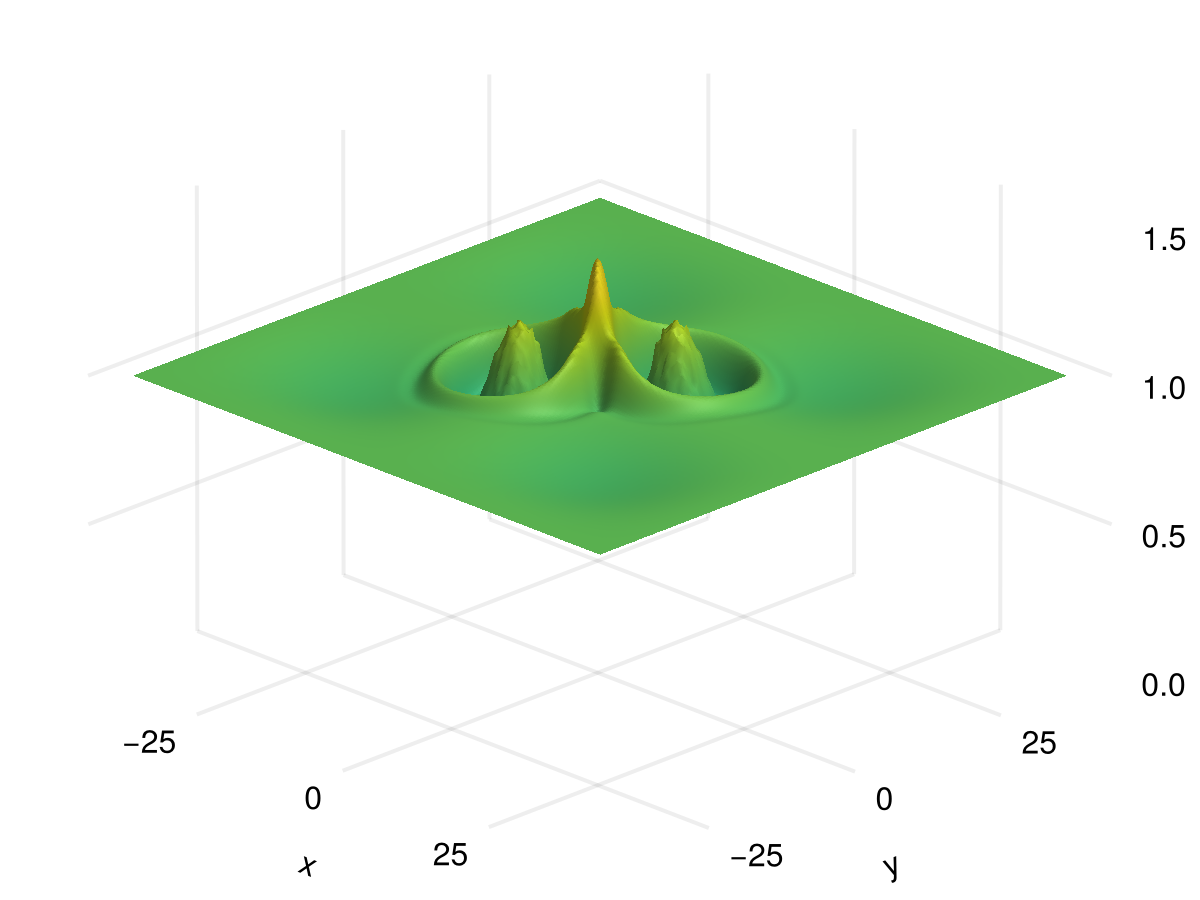}\\
        {\includegraphics[width=\linewidth]{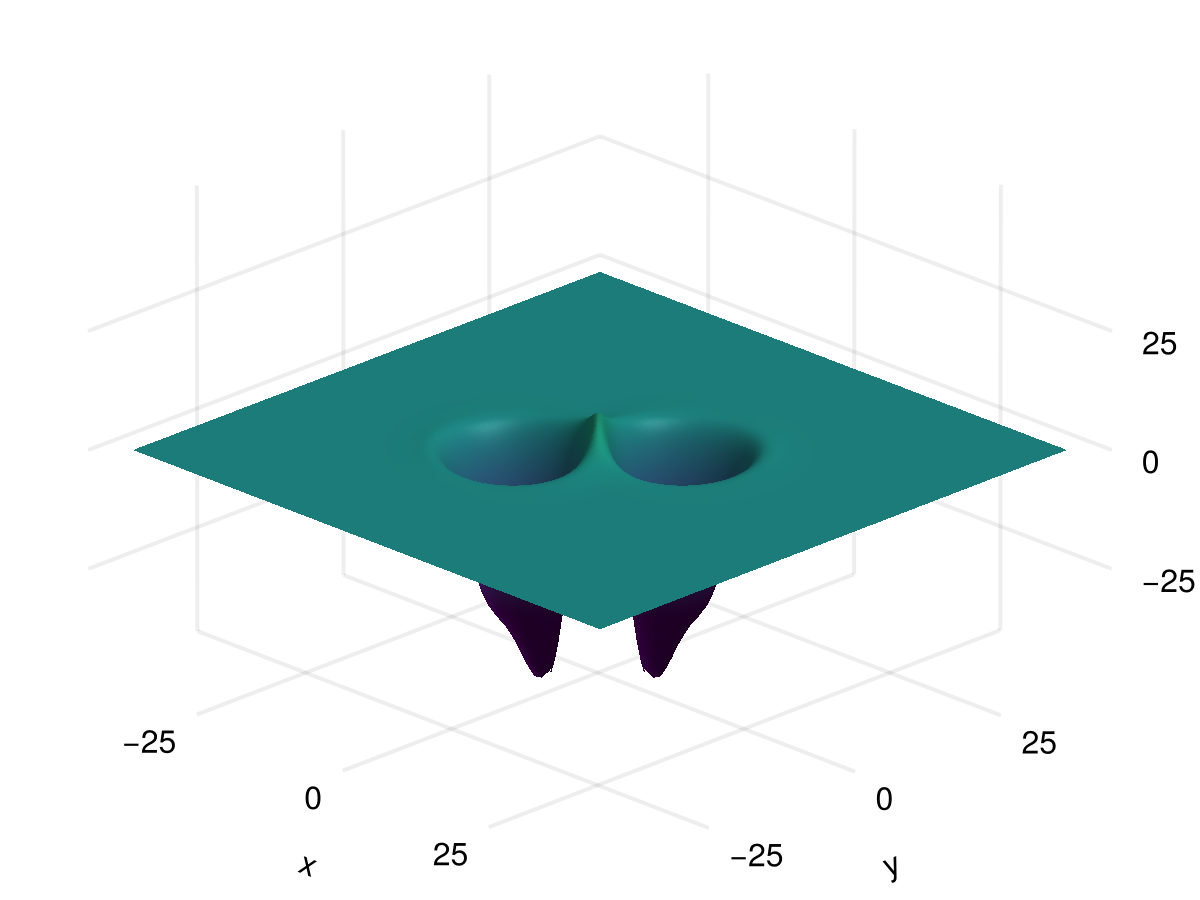}}
    \end{minipage}
    }\hfill
    \subfloat[]{%
    \begin{minipage}{0.315\textwidth}
        \centering
        \includegraphics[width=\linewidth]{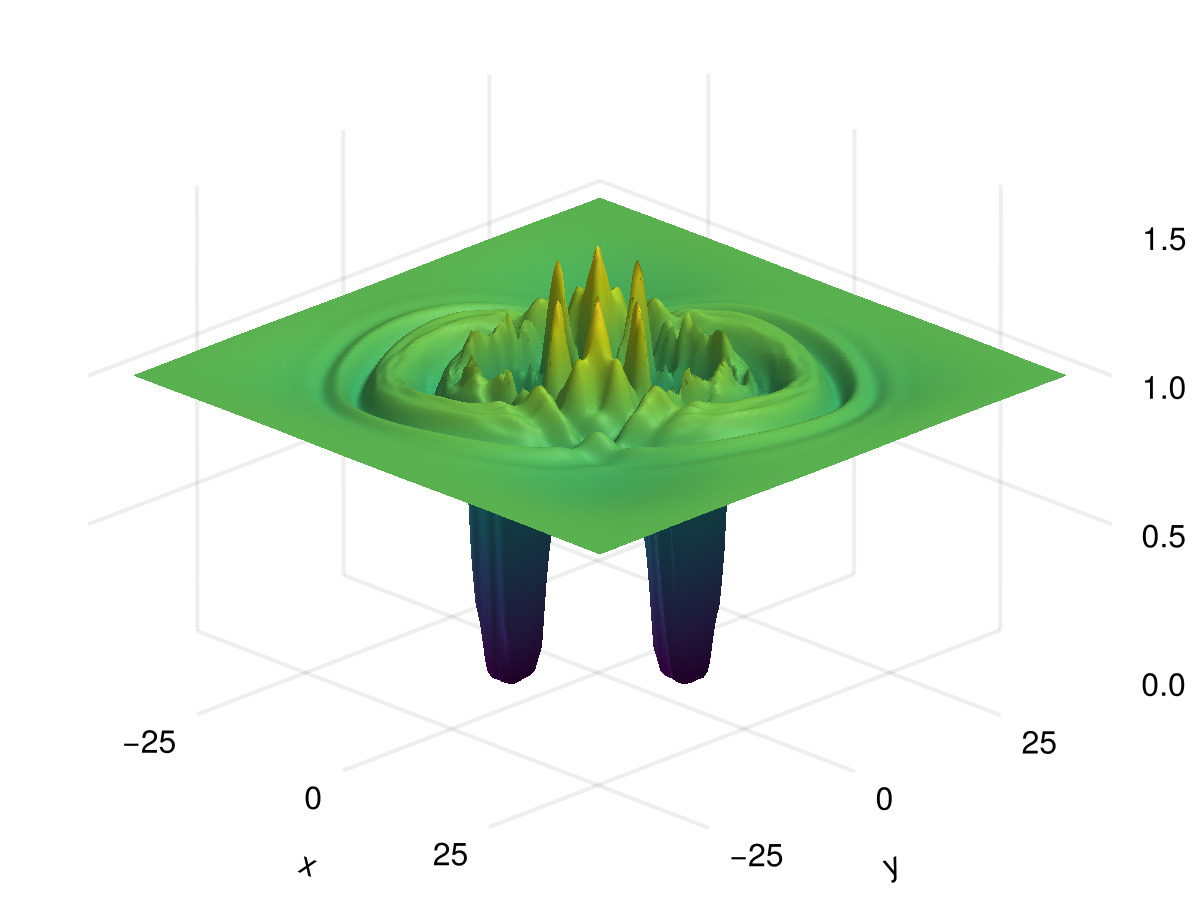}\\
        {\includegraphics[width=\linewidth]{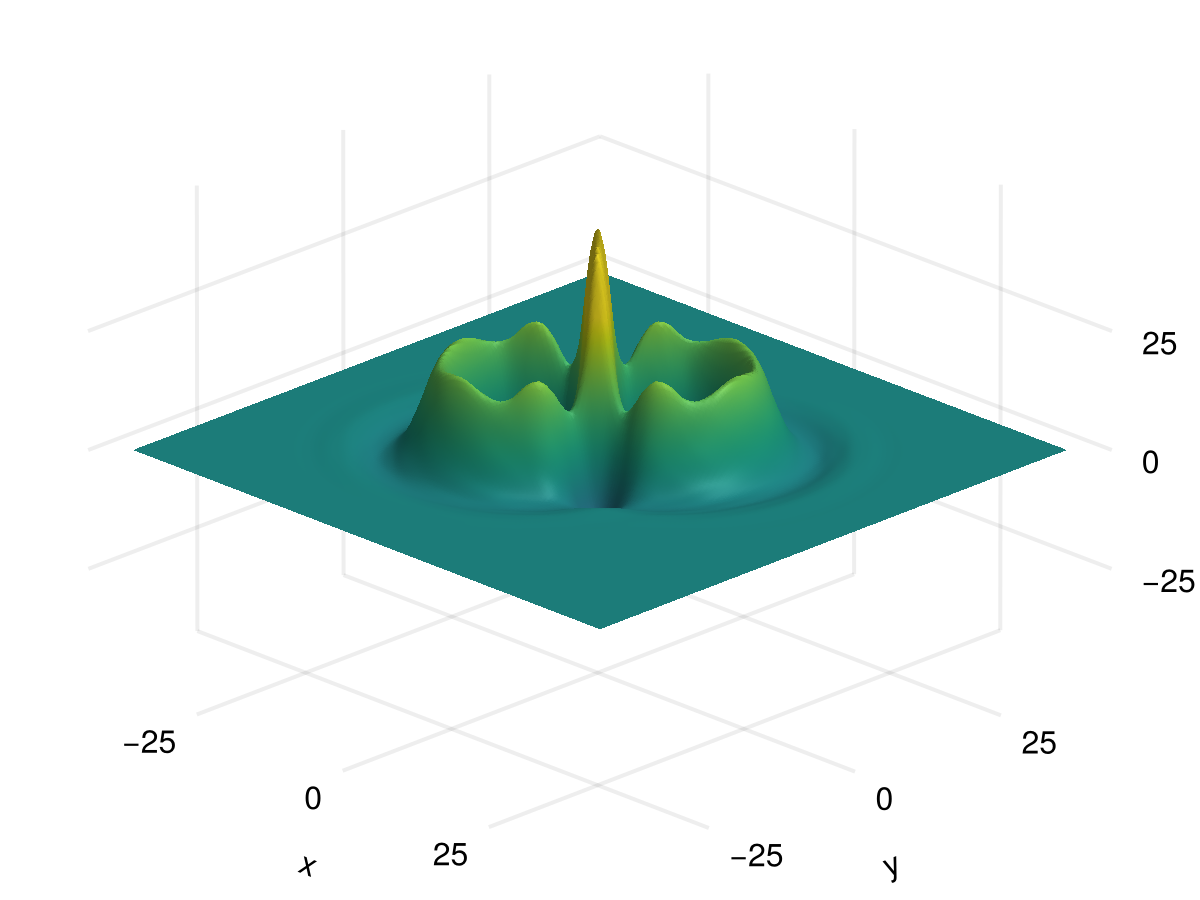}\label{fig:simulation1-vortexFormation}}
    \end{minipage}
    }\hfill
    \subfloat[]{%
    \begin{minipage}{0.35\textwidth}
        \centering
        \includegraphics[width=\linewidth]{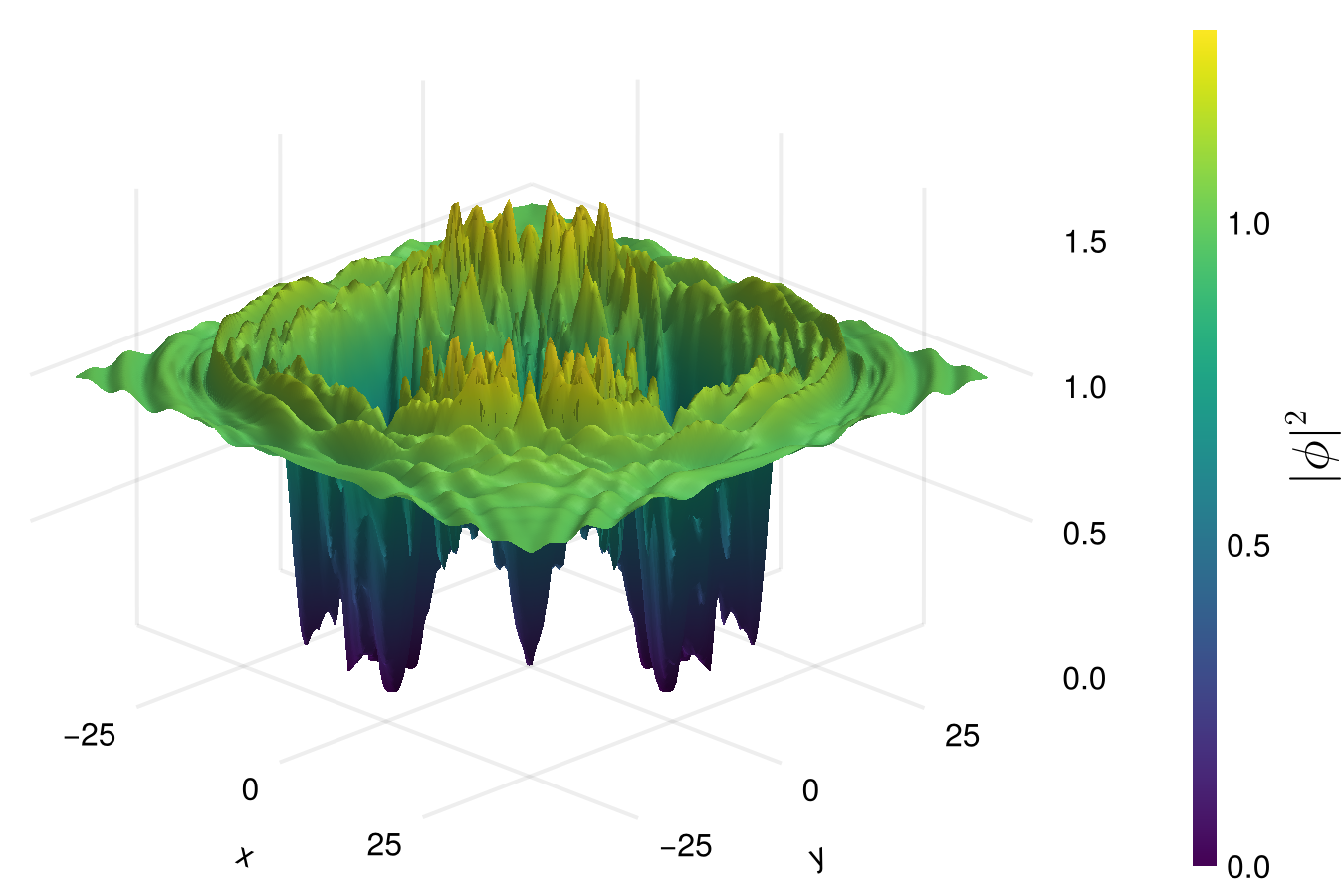}\\
        \includegraphics[width=\linewidth]{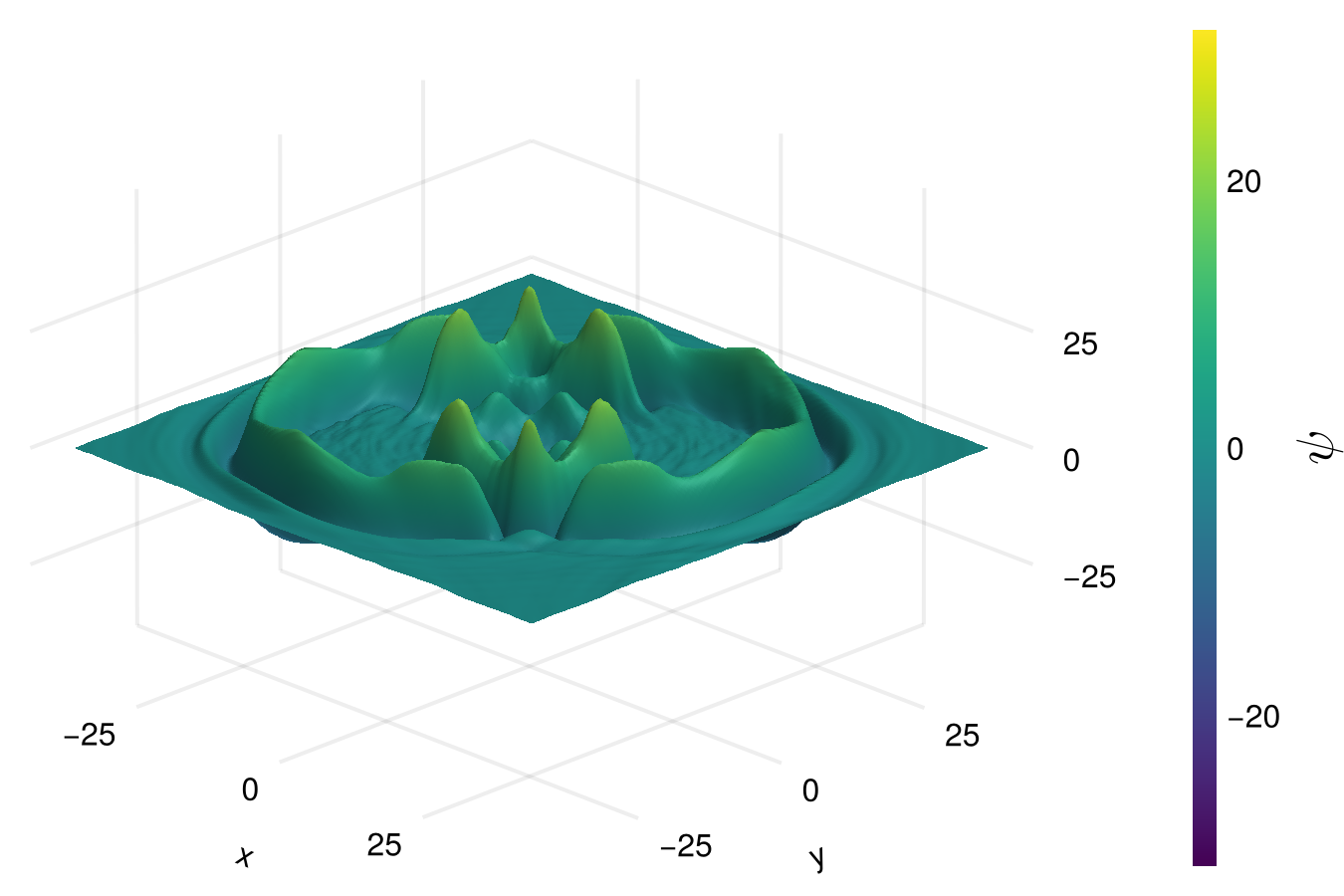}
    \end{minipage}
    }
    \caption{Three snapshots of the time evolutions of fields $\phi$ and $\psi$ (top and bottom rows, respectively) for parameters $A=94$, $v=0.55$ and $k=0.05$.
    Columns show the system at times near peak Gaussian wavepackets overlap, during initial vortex production, and at later times, respectively.}
    \label{fig:simulation1-main}
\end{figure*}

% cases
We present here an example case and the parameter space scans over the parameters amplitude $(A)$, velocity $(v)$, and width $(k^{-1})$ for vortex production.
Fig.~\ref{fig:simulation1-main} shows an example of the time evolution of the $\phi$ and $\psi$ fields in the case of vortex production, with a maximum of 4 vortices (+4 anti-vortices), for parameters $A=94$, $v=0.55$, and $k=0.05$. 
In Fig.~\ref{fig:simulation1-vortexFormation}, the initial vortex formation is shown where $\phi$ vanishes at certain points, which is seen by the troughs colored by deep purple in the plot. Note that vanishing $\phi$ alone does not imply a vortex; vortex locations are identified via the winding number, as discussed in Section~\ref{numerical}. It is noteworthy to mention that the plots clearly illustrate the impact of $\psi$ on $\phi$, despite the absence of an explicit interaction term between them.

\begin{figure*}
    \centering
    \subfloat[]
    {\includegraphics[width=0.46\textwidth]
        {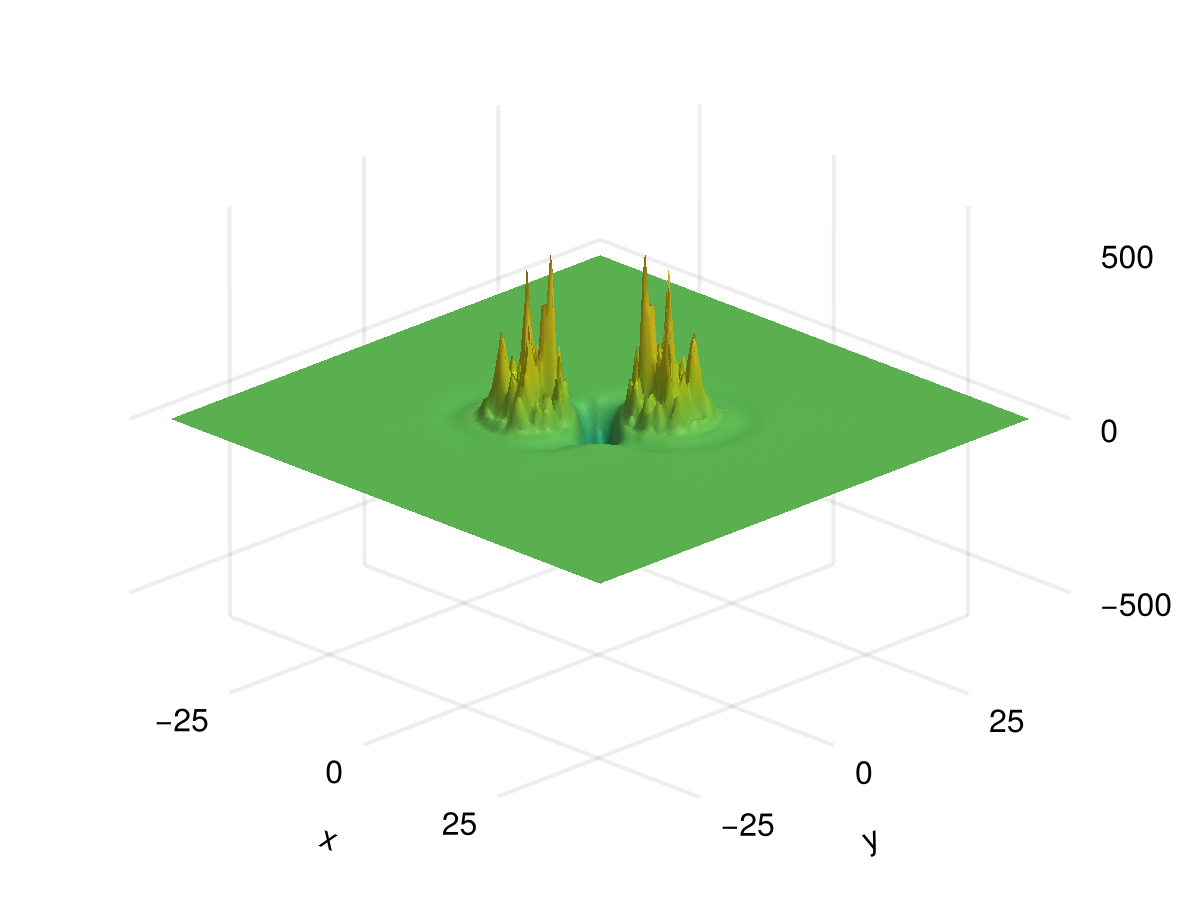}
        \label{fig:s1-ZED-before}}\hfill
    \subfloat[]
    {\includegraphics[width=0.520\textwidth]
        {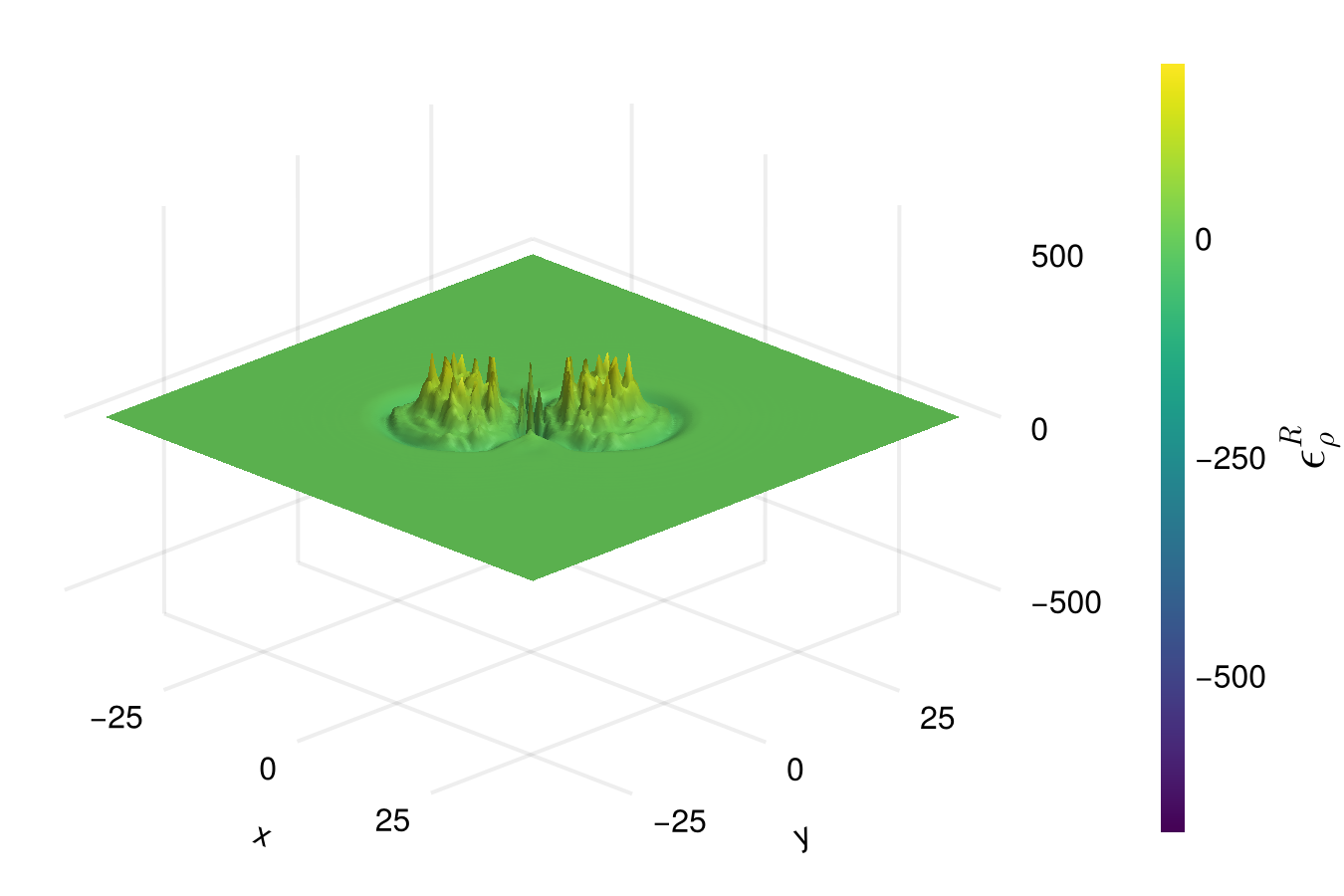}
        \label{fig:s1-ZED-after}}\hfill
    \caption{\label{fig:s1-ZED-main} 
    (a), (b) Energy density of $\rho$ ($\epsilon_{\rho}^R$) before and after the vortex formation respectively for parameters $A=94$, $v=0.55$ and $k=0.05$. The imprints of the $\phi$ and $\psi$ interactions are visible. (a) $\epsilon^{R}_{\rho}$ fluctuates rapidly until the formation of the first vortices, and (b) simmers down after the formation.
}
\end{figure*}

\begin{figure*}
    \centering
    \subfloat[]
    {\includegraphics[width=0.49\textwidth]
        {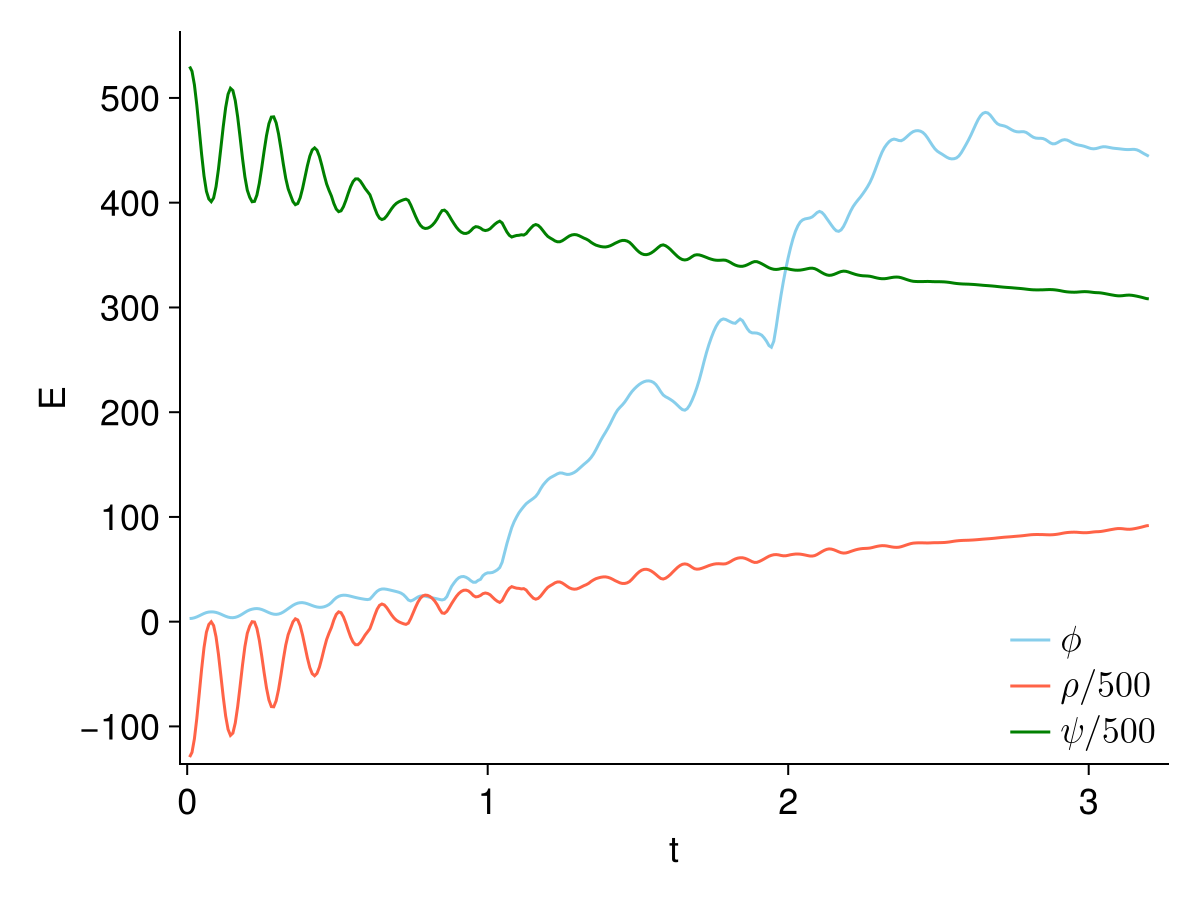}
        \label{fig:s1-indv-energies}}\hfill
    \subfloat[]
    {\includegraphics[width=0.49\textwidth]
        {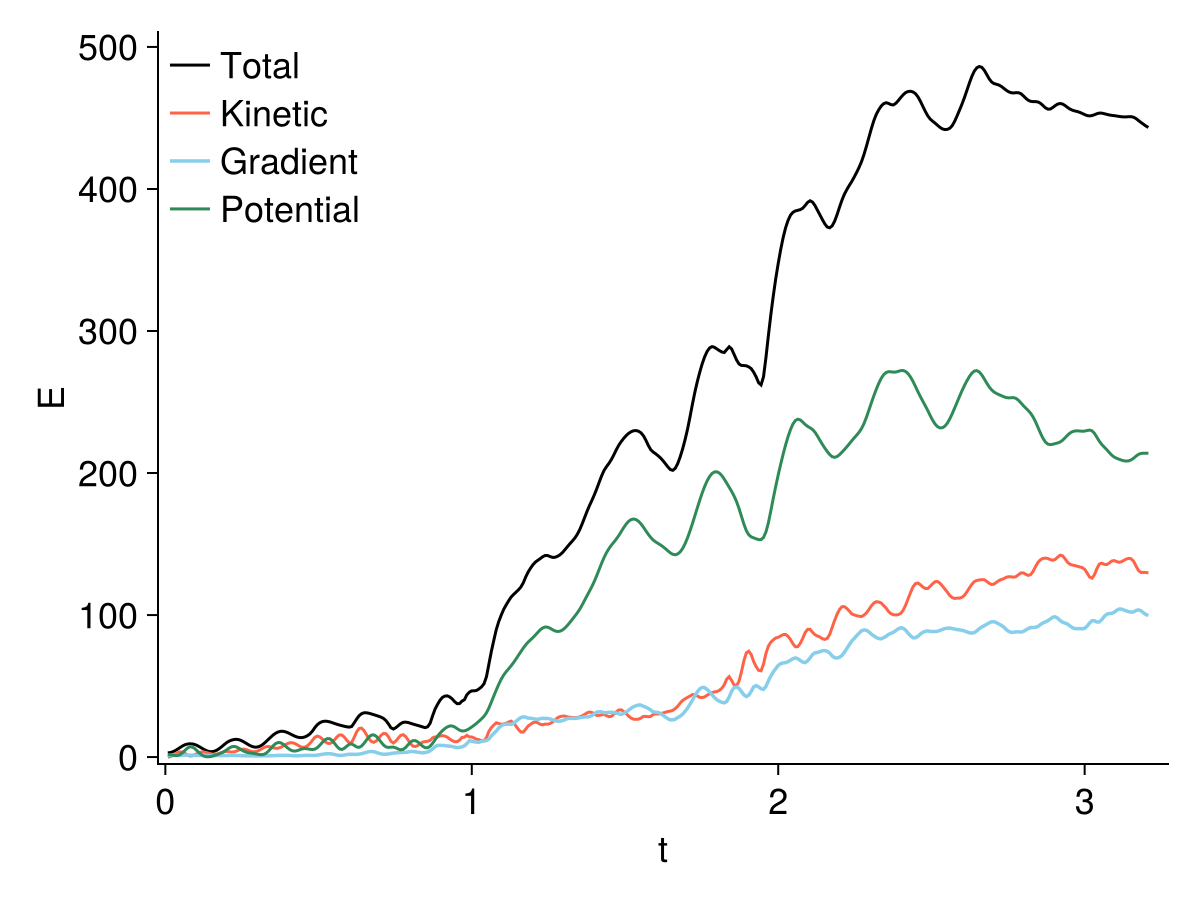}
        \label{fig:s1-phi-energies}}\hfill
    \caption{\label{fig:s1-energies} 
    (a) Total energies of individual fields over time for parameters $A=94$, $v=0.55$, and $k=0.05$.
    The interaction energies are included in $\rho$ after suitable renormalizations.
    The total energies for fields $\phi$ and  $\psi$ include only kinetic, gradient, and potential contributions; interaction energies are not included to avoid double counting. 
    (b) Kinetic, gradient, and potential energy contributions of $\phi$ over time. The potential energy experiences a jump around the time of vortex formation at $t=1$.
}
\end{figure*}

\begin{figure}
    \centering
    {\includegraphics[width=1.0\linewidth]{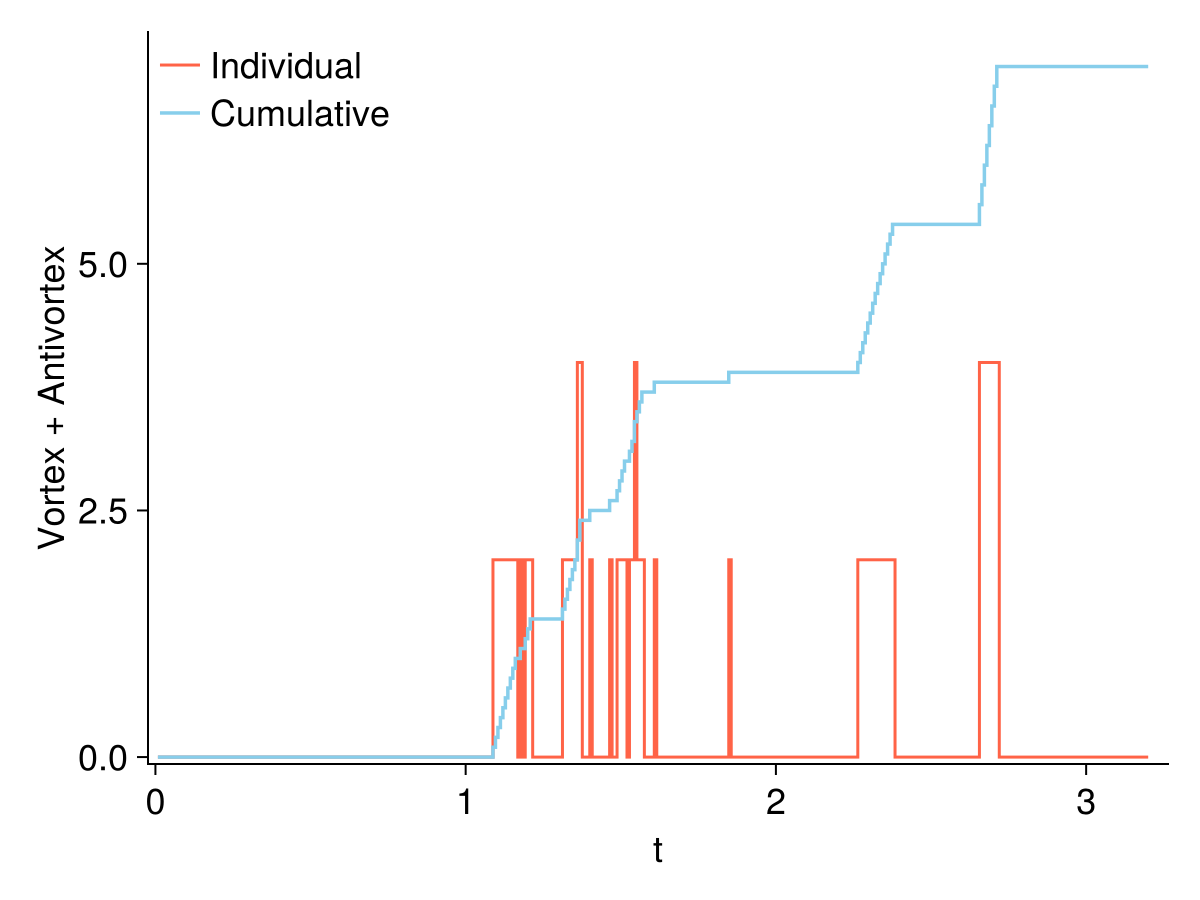}}
    \caption{
        Number of vortices over time (orange) and the cumulative sum over the number of vortices over time (blue) for parameters $A=94$, $v=0.55$, and $k=0.05$. The cumulative curve is rescaled by a constant factor for visual clarity, as only its temporal structure is relevant.
    }
    \label{fig:s1-vortex-vs-time}
\end{figure}

Figures \ref{fig:s1-ZED-before} and \ref{fig:s1-ZED-after} illustrate the energy density $\epsilon_{\rho}^R$ (\ref{eqn:edensity_renorm}) of the field $\rho$. The effects of $\phi$ and $\psi$ on $\epsilon_{\rho}^R$ are clearly visible owing to their interactions.
Initially, since $\phi$ is in its vacuum state, the fluctuations we observe in $\epsilon_{\rho}^R$ during the early stages are solely attributed to the effects of the Gaussian wavepackets. Figure \ref{fig:s1-ZED-before} captures these early times, where we note that the energy density fluctuates rapidly until the initial vortex formation.
As the vortex formation begins, there is a noticeable slowdown in the fluctuations of $\epsilon_{\rho}^R$, as demonstrated in Figure \ref{fig:s1-ZED-after}. This can also be seen in Fig.~\ref{fig:s1-energies} as the oscillation in the total energy of $\rho$ dampens suddenly around $t=1$ when the vortex formation begins. Concurrently, we also observe a sharp increase in the total energy of the field $\phi$ in the form of potential energy, as shown in Figure \ref{fig:s1-phi-energies}. 
Fig.~\ref{fig:s1-energies} also indicates that only a tiny fraction of the initial energy passes from $\psi$ onto the $\phi$ side of the bridge to produce the vortices. 
Energy transfer can, in principle, be enhanced by adjusting the couplings or by adopting more tailored initial conditions for $\psi$; however, to preserve the simplicity of the toy model, these possibilities were not explored. That said, energy transfer is not the only factor in vortex production. As shown in Fig.~\ref{fig:s1-vortex-vs-time}, just before $t=2$ all the initial production ceases, and the generated vortices annihilate each other. Nevertheless, we observe a late production period after $t=2$, well after the initial Gaussian scattering has ended. 
This implies that the vortex production is not merely a direct kinematic debris of the scattering of the wavepackets. Instead, the high initial energy triggers a persistent instability that continues to evolve independently of the original interaction.

\begin{figure}
    \centering
    {\includegraphics[width=0.8\linewidth]{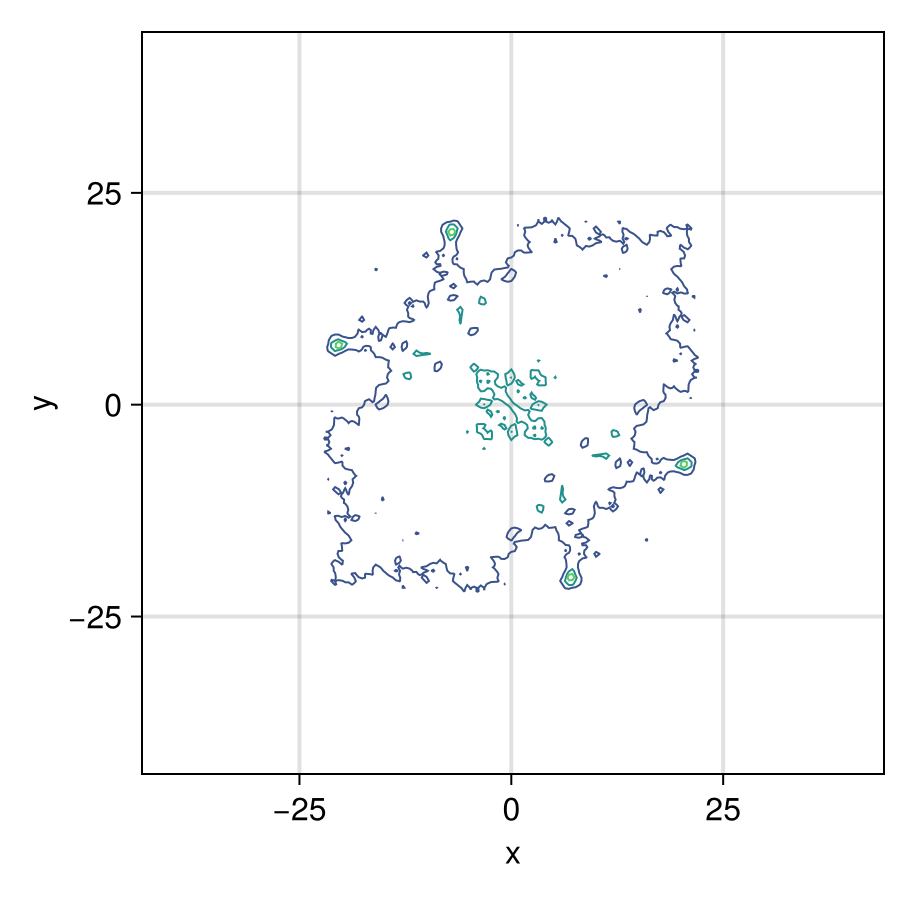}}
    \caption{
        Contour plot of energy density of the quantum field $\rho$ ($\epsilon_{\rho}^R$) for $A=94$, $v=0.55$ and $k=0.05$. The four points of energy clumps on either side of the $y=-x$ line are long-lasting peaks in the energy density that move out of the interaction region in a direction roughly orthogonal to the initial scattering momenta.
    }
    \label{fig:s1-ZED-soliton-like}
\end{figure}

Additionally, we see energy in $\epsilon_{\rho}^R$ that is not directly related to the Gaussian wavepackets or the produced vortices. This energy manifests itself as quantum radiation. This becomes more apparent in the late stages of the simulation, once the Gaussian wavepackets have sufficiently dissipated and the energy density $\epsilon_{\rho}^R$ is no longer dominated by their interaction, revealing the underlying field radiation. Furthermore, in some simulations, peaks in the energy density emerge within this quantum radiation, as seen in Fig.~\ref{fig:s1-ZED-soliton-like}. These structures persist for a considerable time after they are created, but eventually start to disperse towards the end of the simulations.

\begin{figure*}[]
    \centering
    \subfloat[]
    {\includegraphics[width=0.315\textwidth]
        {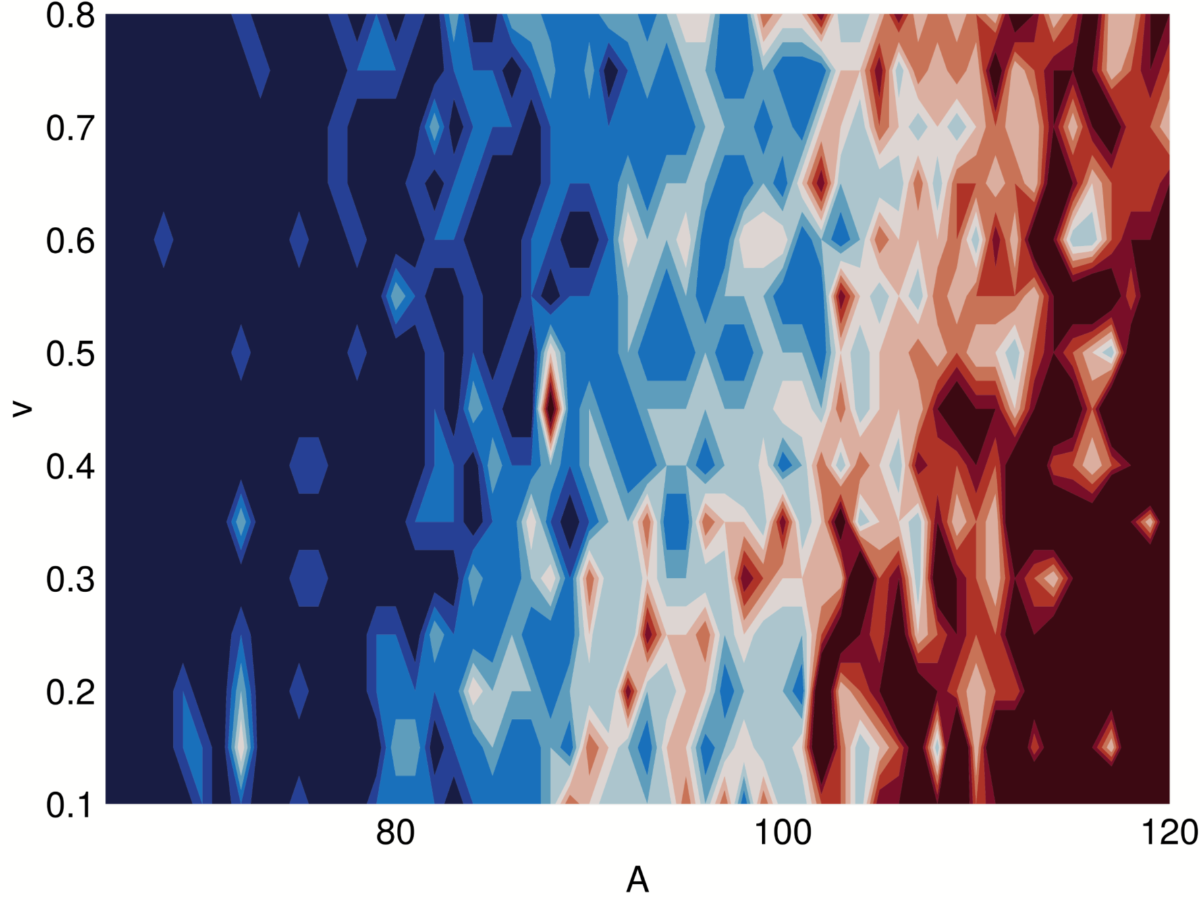}
        \label{fig:A_v-wide}}\hfill
    \subfloat[]
    {\includegraphics[width=0.315\textwidth]
        {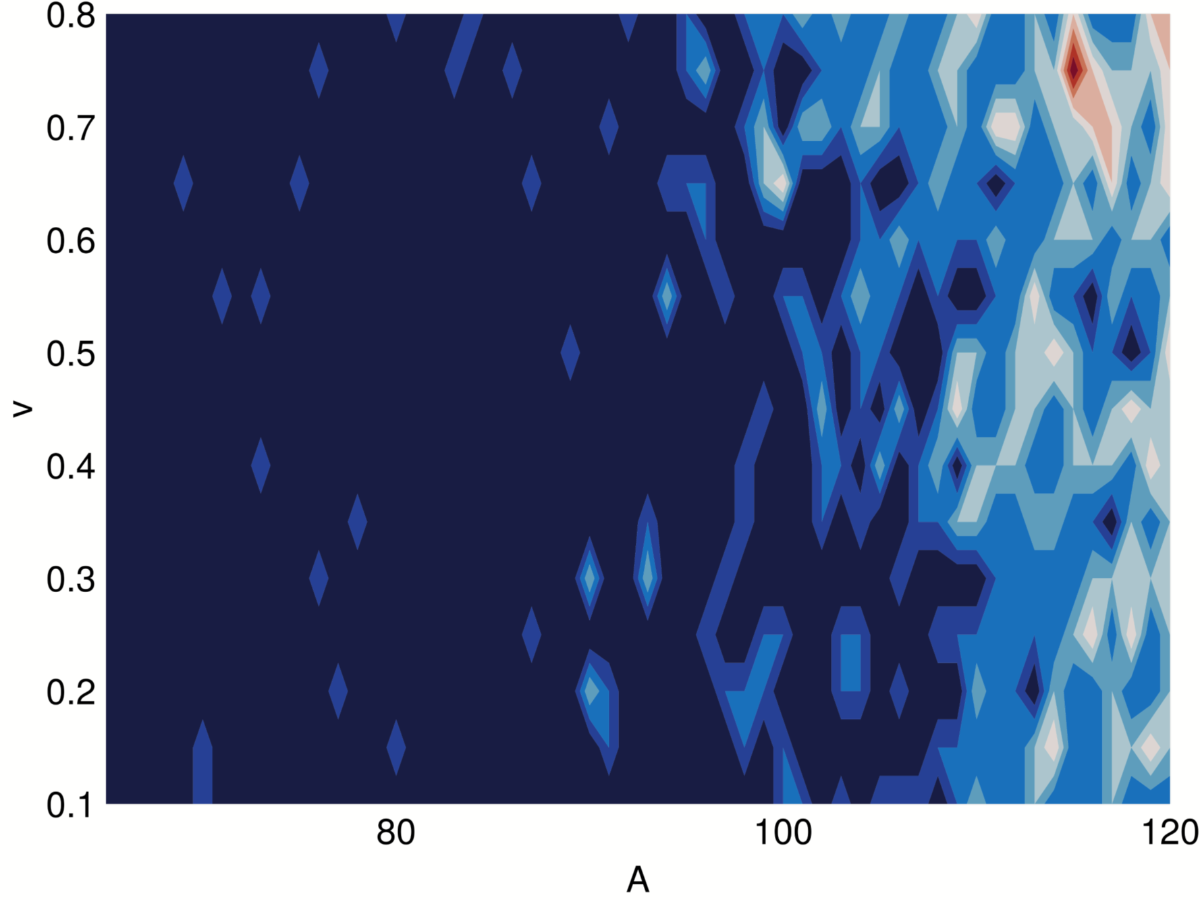}
        \label{fig:A_v-mid}}\hfill
    \subfloat[]
    {\includegraphics[width=0.3425\textwidth]
        {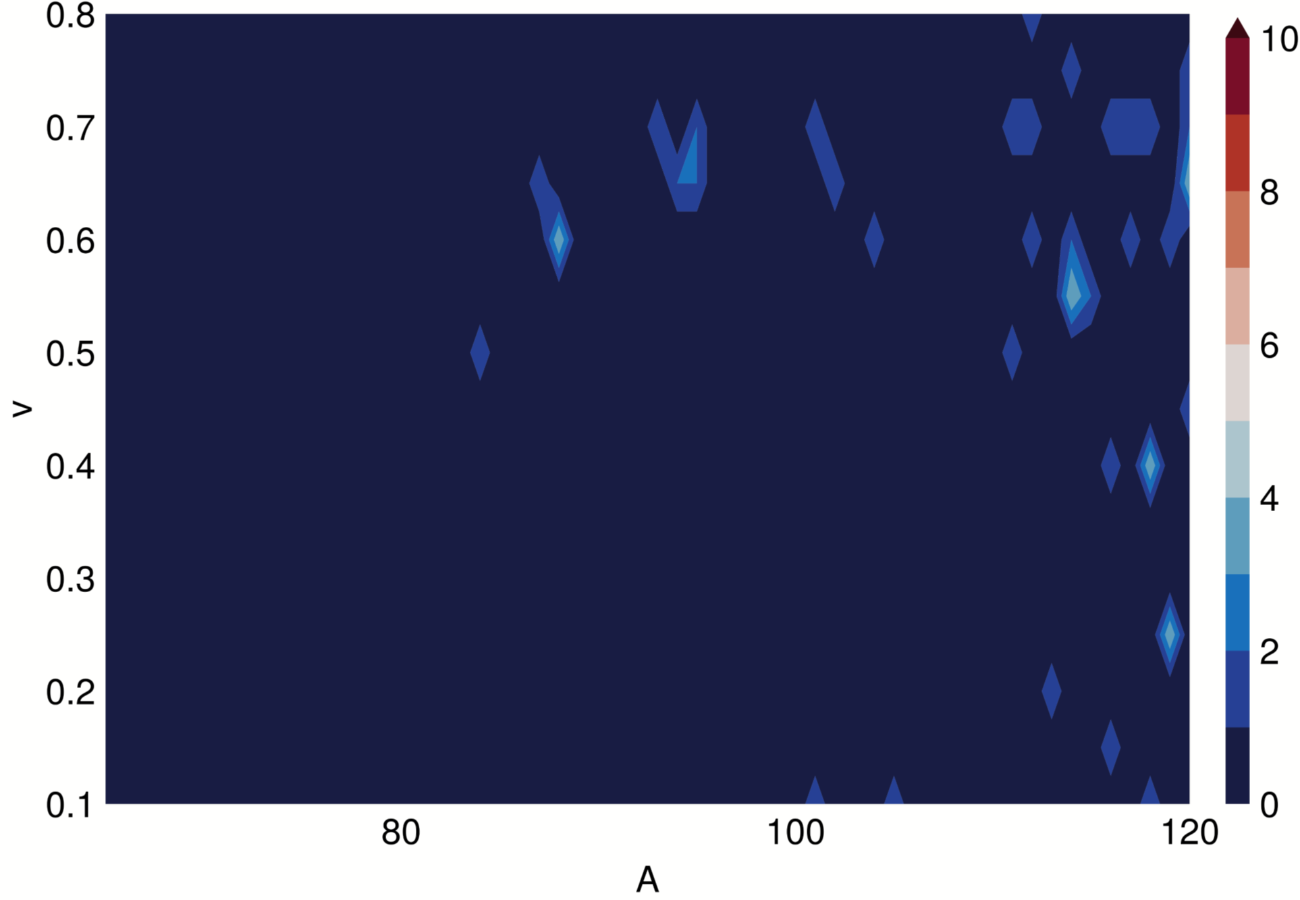}
        \label{fig:A-v_slim}}\hfill
    \caption{\label{fig:A-v_main} 
    Global vortex production in the amplitude ($A$) and velocity $(v)$ plane as a contour plot for (inverse) widths 
     (a) $k=0.03$, (b) $k=0.05$, and (c) $k=0.09$. 
     The color bar shows the maximum number of vortices (winding $=1$) produced during the simulation for the given parameters. These vortices are also accompanied by the same number of anti-vortices (winding =$-1$).
     Although the vortex yield is chaotic in the $A$ and $v$ parameter space, larger incoming amplitude tends to yield more vortices, while higher incoming velocity does not have a clear advantage. 
     In going to larger values of $k$, {\it i.e.} smaller wave packet widths, we see fewer cases of vortex production, implying that  vortex production is more efficient with wider wavepackets. 
}
\end{figure*}

% parameter space scans
The production of vortices depends highly on the initial conditions of the Gaussian wavepackets. To determine the optimal conditions for vortex production, we evolved the system with various initial conditions over a light-crossing time. We record and track the vortices (and anti-vortices) produced during the simulation and catalog the simulations by their maximum number of vortices created, as mentioned above. Fig.~\ref{fig:A-v_main} shows the scans of the parameter space of amplitude $A$ and velocity $v$ for different sizes of the Gaussian wavepackets. From the outlook, the wavepacket widths appear to affect the overall trend in vortex production. Fig.~\ref{fig:A-v_main} demonstrates that increasing the initial wavepacket width leads to more vortex production. 

The parameter scans of each width, however, show that the production is highly fractured on the $A-v$ space. There are many holes, even at very high energies, where production would be expected based solely on energy considerations, and many isolated high-production regions. The general tendency appears to be that higher amplitudes and moderate speeds favor vortex production.

\section{Conclusions}
\label{conclusions}

As mentioned in the introduction, the aim of this investigation is to study the transition from the particle sector to the soliton sector when interactions between the two are mediated by quantum degrees of freedom. The setup is relevant to situations such as magnetic monopole creation by light on light scattering. Here, we considered global vortex production in 2+1 dimensions in the scattering of wave packets of a scalar field.
Although the setup is similar to the production of kinks in 1+1 dimensions~\cite{Albayrak:2023dul}, there are some crucial differences. Chief among them is that, if the vortex field $\phi$ is set to zero initially, then the scattering process can only excite $\phi$ in the radial direction, {\it i.e.} the outcome of the scattering does not excite the phase of $\phi$. Since vortices occur due to topological windings of $\phi$, a fixed phase of $\phi$ means that no vortices can be produced. However, the situation is unstable and phase excitations and vortex production are possible if there is a slight deviation of $\phi$ from zero in the initial conditions. In future work, we plan to examine vortex production in a model in which the phase gets excited even if we set $\phi=0$ initially.

Whether vortices are produced depends on the scattering parameters. Our results are best summearized in Fig.~\ref{fig:A-v_main}. where we show the number of vortices produced as we vary the amplitude and velocity of the incoming wave packets. It is clear from these plots that increasing the amplitude enhances vortex production. The dependence on the velocity is less clear. What is clear, however, is that vortex creation is a chaotic process. This may be a hint that the physical mechanism behind soliton production may be due to a resonance between different degrees of freedom, similar to the resonance that has been examined in the context of kink-antikink 
scattering~\cite{RoyH:2006bwi}.

\acknowledgements
We thank Gil Speyer for assistance with access to computational resources and technical support.
The computations were performed on the Sol supercomputer at Arizona State University \cite{HPC:ASU23} and on the Launch cluster at Texas A\&M University.
This work was supported by the U.S. Department of Energy, Office of High Energy Physics, under Award No.~DE-SC0019470.

\bibstyle{aps}
\bibliography{paper}

\end{document}